\documentclass[aps,prb,showpacs,twocolumn]{revtex4}
\usepackage{graphicx}
\begin{document}
\newcommand{\beq}{\begin{equation}}
\newcommand{\eeq}{\end{equation}}
\newcommand{\bear}{\begin{eqnarray}}
\newcommand{\eear}{\end{eqnarray}}
\renewcommand{\a}{\alpha}
\newcommand{\ta}{\tilde{\a}}
\renewcommand{\b}{\beta}
\newcommand{\tb}{\tilde{\beta}}
\newcommand{\rd}{\mathrm{d}}
\newcommand{\ee}{\mathrm{e}}
\newcommand{\lp}{\left}
\newcommand{\rp}{\right}
\newcommand{\bs}{\mathbf}
\newcommand{\bz}{\bs{\hat{z}}}
\newcommand{\bx}{\bs{\hat{x}}}
\newcommand{\by}{\bs{\hat{y}}}
\newcommand{\hb}{\hbar}
\newcommand{\g}{\gamma}
\newcommand{\e}{\epsilon}
\newcommand{\pd}{\partial}
\renewcommand{\l}{\lambda}
\renewcommand{\d}{\delta}
\renewcommand{\k}{\kappa}
\renewcommand{\o}{\omega}
\renewcommand{\O}{\Omega}
\newcommand{\ofm}{\o_{fm}}
\newcommand{\omp}{\o_{m\, \parallel}}
\newcommand{\Omp}{\O_{m\, \parallel}}
\newcommand{\qmp}{q_{m\, \parallel}}
\newcommand{\ohp}{\o_{h\, \parallel}}
\newcommand{\Ohp}{\O_{h\, \parallel}}
\newcommand{\kmp}{k_{m\, \parallel}}
\newcommand{\s}{\sigma}
\renewcommand{\bm}{\bs{M}}
\newcommand{\sm}{\bs{m}}
\newcommand{\bmo}{\bs{M}_0}
\newcommand{\bmp}{\bs{M_\perp}}
\newcommand{\smp}{\bs{m_\perp}}
\newcommand{\bmx}{\bs{M_x}}
\newcommand{\bmy}{\bs{M_y}}
\newcommand{\smx}{\bs{m_x}}
\newcommand{\smy}{\bs{m_y}}
\newcommand{\nb}{\nabla}
\newcommand{\tm}{\times}
\newcommand{\bb}{\bs{B}}
\newcommand{\bbo}{\bs{B}_0}
\renewcommand{\sb}{\bs{b}}
\newcommand{\bbz}{\bs{B_z}}
\newcommand{\ba}{\bs{A}}
\newcommand{\bk}{\bs{k}}
\newcommand{\bh}{\bs{H}}
\newcommand{\bho}{\bs{H}_0}
\newcommand{\sjs}{\bs{j}_s}
\newcommand{\su}{\bs{u}}
\newcommand{\sh}{\bs{h}}
\newcommand{\z}{\zeta}
\newcommand{\Ft}{\tilde{F}}
\newcommand{\Gt}{\tilde{G}}
\newcommand{\mxo}{M_{x\,1}}
\newcommand{\mxd}{M_{x\,2}}
\newcommand{\mxt}{M_{x\,3}}
\newcommand{\myo}{M_{y\,1}}
\newcommand{\myd}{M_{y\,2}}
\newcommand{\myt}{M_{y\,3}}
\newcommand{\mxi}{M_{x\,i}}
\newcommand{\myi}{M_{y\,i}}
\newcommand{\mxy}{M_{x,y}}
\newcommand{\oth}{\o_{th}}
\newcommand{\Oth}{\O_{th}}
\newcommand{\lt}{\tilde \l}
\newcommand{\se}{\bs{e}}
\newcommand{\pp}{{(+)}}
\newcommand{\mm}{{(-)}}
\newcommand{\ppm}{{(\pm)}}
\newcommand{\mpm}{m^{(\pm)}}
\newcommand{\bpm}{b^{(\pm)}}
\newcommand{\hpm}{h^{(\pm)}}
\newcommand{\epm}{e^{(\pm)}}
\newcommand{\zpm}{\z^{(\pm)}}
\newcommand{\zr}{\z_{res}^{(+)}}
\newcommand{\im}{\mathrm{Im}}
\newcommand{\re}{\mathrm{Re}}
\newcommand{\parl}{\parallel}
\newcommand{\sgn}{\mathrm{sign}}
\newcommand{\ibid}{{\it ibid }}
\newcommand{\etl}{{\it et al.}}
\setlength\arraycolsep{1pt}

\title{Microwave response and spin waves in superconducting ferromagnets}
\author{V.~Braude}
\affiliation{Racah Institute of Physics, The Hebrew University of
Jerusalem, Jerusalem 91904, Israel}
\affiliation{Kavli Institute
of Nanoscience,  Delft University of Technology,  2628 CJ Delft,
The Netherlands}
\date{\today }
\begin{abstract}
Excitation of spin waves is considered in a  superconducting
ferromagnetic slab with the equilibrium magnetization both
perpendicular and parallel to the surface. The surface impedance
is calculated, and its behavior near propagation thresholds is
analyzed. The influence of nonzero magnetic induction at the
surface is considered in various cases. The results provide a
basis for the investigation
 of materials with coexisting superconductivity and
magnetism by microwave
response measurements.
\end{abstract}
\maketitle
\section{Introduction}
The problem of the coexistence of superconductivity and magnetism
has attracted renewed interest during the last decade due to the
discovery of unconventional and high-$T_c$ superconductors, in
which such a coexistence seems to be realized.  The presence of
magnetic properties  has been supported by experimental evidence
in such materials as high-$T_c$ ruthenocuprates, \cite{felner}
Sr$_2$RuO$_4$, \cite{luke} ZrZn$_2$, \cite{Pfleiderer} and
UGe$_2$. \cite{Saxena}

Experimental investigation of ferromagnetism coexisting with
superconductivity is hampered by the fact that the spontaneous
magnetic moment is screened by the Meissner currents in
macroscopic samples. Therefore the usual methods, such as the
Knight shift or muon spin relaxation, are only able to detect
small remanent fields near inhomogeneities, defects etc. On the
other hand, dynamical measurements - e.~g.,~ excitation of spin
waves - provide a direct probe of bulk magnetization and are
sensitive not only to its magnitude, but also the direction and
degree of uniformity. There have been reports in the literature
\cite{tallon} of microwave measurements in powder or ceramic
polycrystalline samples made of materials with coexisting
superconductivity and magnetism (SCFM).

 Theoretical research on macroscopic
electromagnetic properties of SCFM's was initiated by Ng and
Varma, \cite{Ng, Ng2} who have considered the bulk magnetic
response and calculated the spin-wave modes for the spontaneous
vortex phase. This regime, in which vortices are created in the
sample without any applied magnetic field, is realized when the
spontaneous magnetization is strong enough - namely, larger than
$H_{c 1}/4\pi$, with $H_{c 1}$ being the first critical field.
However,
 in order to make direct
comparison with microwave experiments, the finite-sample response
- namely, the surface impedance - is needed. Unlike in usual
metals, dielectrics, or paramagnetics, this quantity cannot be
trivially related to the bulk properties in SCFM's, but requires a
separate study. Hence the next step was done by the present author
and Sonin, \cite{BS,SW,pwave} who have considered the
electromagnetic response of a thick slab, concentrating on the
Meissner regime, in which static magnetic fields are screened in
the bulk. In this regime, static magnetic measurements are
difficult, and hence microwave experiments may be of particular
interest.
 Spin-wave excitation and propagation were  considered for a SCFM
slab  with the magnetization perpendicular to the surface
(perpendicular geometry). \cite{BS} While the analysis was
initially carried for the case of Landau-Lifshitz (LL) SCFM's with
spin magnetism, the results have been generalized to the case of
$p$-wave superconductors with orbital magnetism. \cite{pwave} It
was shown that these materials are in general described by more
complicated magnetic dynamics than LL dynamics. However, two
specific cases - namely, when the waves propagate either parallel
or perpendicular to the spontaneous magnetization direction - can,
in fact, be described by LL dynamics, though corresponding to
different values of the magnetization, thus demonstrating the
relevance of the analysis to unconventional superconductors with
orbital magnetism. Also, it was found that SCFM's can support a
surface wave, \cite{SW} and its spectrum was analyzed as
 compared to  normal ferromagnets.

In the present work, I extend the microwave response analysis
started in Ref.~ \onlinecite{BS}, including both the perpendicular
and parallel geometries (when the magnetization is perpendicular
or parallel to the surface). Noteworthy, the response is shown to
exhibit a resonance related to excitation of the surface wave. The
influence of applied magnetic fields is also considered, leading
to completely different effects in the perpendicular and parallel
cases. In the former case, the field penetrates the sample,
pushing it into the vortex state. In the latter case, on the other
hand, the field is screened inside the Meissner layer and the
system loses its uniformity.

\section{The model}
We consider a SCFM in form of a slab which is thick enough so that
it is possible to neglect  reflection of waves from the second
boundary. The material is assumed to possess a ferromagnetic
anisotropy of the easy axis type, with the direction of the easy
axis being chosen as the $\bz$ direction, so that the equilibrium
magnetization is $\bs{M}_0=M_0 \,\bz$. For $4\pi M_0<H_{c 1}$,
which we assume,
the material is in the Meissner phase, except when an external
magnetic field is applied perpendicular to the sample surface (in
which case the material is in the vortex state). The free energy
functional for the  material is given by \cite{Ng,BS}
\bear
  F=\int d^3 x \lp[ \frac{\a}{2} M_{x,y}^2+\frac{\g}{2}(\pd_i M_j)(\pd_i M_j)
  \rp.
\nonumber \\
  + \lp.
  \frac{1}{8 \pi \l^2}\lp(\nabla \phi
 \frac{\Phi_0}{2\pi}-\bs{A}\rp)^2+\frac{B^2}{8\pi}
-\bs{B\cdot M}
\rp], \eear where $\bm$ is the magnetization, $\bm_{x,y}$ is the
magnetization component normal to the easy axis, $\pd_i=\pd/\pd
x_i$, $\a>0$ is the anisotropy parameter (note that it has a
different normalization here than in Ref.~\onlinecite{BS}), $\phi$
is the phase of the superconducting order parameter, $\Phi_0 = h
c/2 e$ is the magnetic flux quantum, $\ba$ is the vector
potential, and $\bb=\nabla \tm \ba$ is the magnetic induction. The
applied static magnetic field is parallel to $\bz$ and so is the
static component of the magnetic induction $\bbo$. The parameter
$\g$ characterizes the stiffness of the spin system. The
anisotropy is assumed to be strong enough - namely, $\d\a =
\alpha-4\pi>0$ - so that the system is stable (or at least
metastable) against flips of the equilibrium magnetization $\bmo$.
In accordance with the micromagnetism approach, the absolute value
of $\bm$ is assumed to be constant, and hence terms dependent on
$|\bm|$ were omitted. We consider a single-domain configuration
since this configuration is either stable \cite{krey} (for strong
enough superconductivity $2\pi \l<\g \a^{1/2}$) or at least
metastable (in the opposite case) with respect to domain
formation, due to the anisotropy, as explained above. In our
treatment, the dissipation initially is not taken into account
explicitly, so its only manifestation is the absence of reflected
waves from the other surface of the sample. Then, having obtained
the result, we generalize it,
including explicitly Ohmic normal currents.

The spin dynamics is governed by the Landau-Lifshitz equation \cite{akhi}
\beq
  \frac{\rd \bs{M}}{\rd t}=-g\lp(\bs{M}\times \frac{\d F}{\d \bs{M}} \rp)
 \label{LL}
 \eeq
where $g$ is the gyromagnetic factor. Since we are concerned here with motion
near the equilibrium, we can decompose
$  \bm=\bmo+ \sm$ and $\bb=\bbo+\sb$,
where $\sm \perp \bmo$ and $\sb$ are dynamical parts. For a
given frequency
Eq.~(\ref{LL}) yields
 \beq \label{eq:dm} -i\omega \, \sm=- g\,
\bmo \tm \lp\{(\a-\g^2 \nb^2)\sm- \sb\rp\}- g\, \bbo \tm \sm. \eeq

Another equation, needed for the dynamical part of the magnetic induction
$\sb$,
 is the London equation
  \beq \label{eq:london}
   \lp(1-\nb^2 \l^{2} \rp)
  \sb=4\pi \l^{2}\, \nb \tm \nb \tm \sm~.
\eeq
The same equation also determines the decay of the static component of the
magnetic induction parallel to the surface inside the sample:
\beq
  \lp(1-\nb^2 \l^{2} \rp)\bbo=0.
\eeq In the regions where $\bbo$ is significant, translational
symmetry is broken, so the resulting equations are quite
complicated. However, if for some reason $\bbo=0$ (e.g. in the
bulk of the sample, where $\bbo$ has already decayed, or by
application of an appropriate external field), the translational
invariance is restored, and the solutions are plane waves $\sim
e^{i(\bk \cdot
 \bs{x}-\o t)}$. Then
the equation of motion for the magnetization is
 \beq
\label{eq:B0=0}
  i \O \,\sm=\bz \tm \lp[ (\a+\g^2 k^2)\sm-
  \frac{4\pi k^2\l^2}{1+k^2\l^2} \smp \rp],
\eeq where $\O = \o/g M_0$ is the frequency in magnetic units and
$\smp$ is the divergence-free component of $\sm$. Note that
retardation effects can be safely neglected here, since the
displacement currents are negligible in comparison with the
superconducting currents as long as the frequency is much smaller
than the plasma frequency, which we, of course, assume. This is in
contrast to insulating ferromagnets, where mixing of spin and
electromagnetic modes can be important. To proceed further, we
will consider two geometries, in which the easy magnetization axis
is perpendicular and parallel to the sample surface.

\section{Perpendicular geometry}
\begin{figure}
\begin{center}
\includegraphics[width=0.4\textwidth]{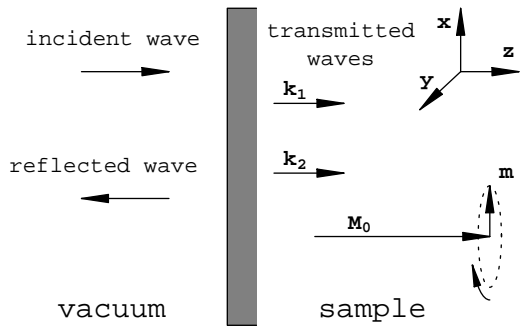}
\end{center}
\caption{The setup in the perpendicular geometry.}
\label{perp}
\end{figure}

Here we consider a case when the easy axis is  perpendicular to
the sample surface, shown in Fig.~\ref{perp}. In this geometry all
quantities vary along the $\bz$ axis, so $\sm$ is divergenceless:
$\smp=\sm$. At zero applied magnetic field the static induction
$\bbo$ vanishes and the system is translationally invariant.
 We first consider this simpler case  and then generalize the results
for the case of nonzero applied magnetic field.

\subsection{Zero applied field}
\begin{figure}
\begin{center}
\includegraphics[width=0.5\textwidth]{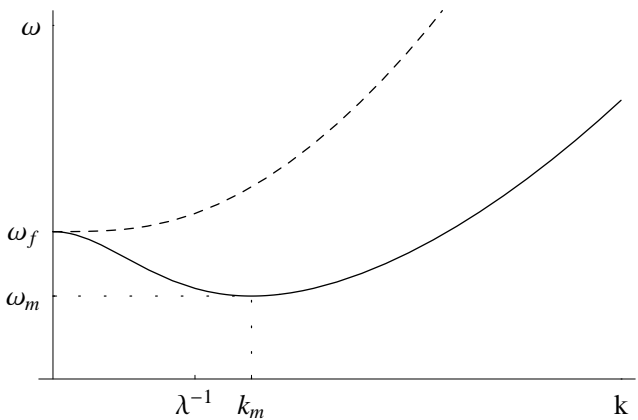}
\end{center}
\caption{The spin-wave spectrum in the perpendicular geometry in
two regimes: stiff, $\e=\sqrt{4\pi}+0.2$, $\a=4\pi+50$ (dashed
line) and soft, $\e=\sqrt{4\pi}-2.35$,  $\a=4\pi+0.013$ (solid
line).} \label{spectr}
\end{figure}

In this regime, $\bbo=0$ and $\smp=\sm$, so
from Eq.~(\ref{eq:B0=0}) the spectrum  of spin waves is
\beq \label{eq:dispperp}
 \O=\pm
  \left(\a+\g^2 k^2- \frac{4\pi k^2\l^2}{1+k^2\l^2} \right)~.
\eeq
The spin-wave modes are circularly polarized, and the two signs
above correspond to two senses of the polarization: $\sm^\ppm=
m^\ppm(\bx \mp i\by)$.
The polarization is determined by the incident electromagnetic
radiation due to continuity of the fields across the slab surface.
Only  positively polarized waves
can propagate inside the sample, and further on we mainly focus on
this polarization. Depending on  the ratio $\e= \g/\l$, the
spectrum can take two different forms, as shown in
Fig.~\ref{spectr}: if $\e>\sqrt{4\pi}$ (high stiffness, strong
superconductivity), the spectrum has a minimum (threshold for spin
wave propagation)  $\O_f=\a$ situated at $k=0$. In the opposite
case, the minimum frequency is at a finite wave vector $k_m=
\l^{-1}(\sqrt{4\pi}/\e-1)^{1/2}$ and  has a lower value
$\O_m=\a-(\sqrt{4\pi}-\e)^2$. In this regime,  waves with wave
vectors satisfying $|k|<k_m$ have  negative group velocity
$d\o(k)/dk$. In order to carry the energy away from the boundary,
these waves should have wave vectors directed toward the boundary.

For given frequency and polarization, two spin-wave modes with
different wave vectors are excited by the incident radiation. For
each mode, the corresponding electric and magnetic fields inside
the sample are found from the Maxwell equation $\partial
\sb/\partial t=-c \mathbf{\nabla} \times \se $ and the London
equation (\ref{eq:london}):
\bear \label{eq:he}
  \sh&=&\sb-4\pi \sm=-\frac{4\pi}{1+k^2 \l^2}\sm~, \nonumber \\
  \se&=&\frac{\o}{k c} \sb \tm \bz = \frac{\o\l }{c}\frac{4 \pi k \l}{1+
   k^2 \l^2} \sm \tm \bz ~.
\eear
The total fields are given by superposition of different modes.
In order to find a
proper superposition of the two spin-wave modes
inside the sample,  an extra boundary condition is needed, in addition to
the usual continuity of  tangential components of the  electromagnetic field
at the sample surface.
This additional condition should be
imposed  on the magnetization. The simplest possible
boundary condition \cite{akhi} is $\pd \sm(z)/\pd z=0$ on the surface,
which physically means
absence of spin currents through the sample surface. In the Fourier space
this condition gives
\beq \label{eq:perpBC}
  k_1 \sm_1+k_2 \sm_2=0~.
\eeq
This equation together with Eq. (\ref{eq:he}) formulates the boundary problem, from which
all fields in the slab can be found.

Now we turn to  calculation of the microwave response, which is
quantitatively expressed by the surface impedance and whose real
part is proportional to the energy absorption by the system.
\cite{LL6} For circularly polarized fields, it is given by \beq
\label{eq:zpm}
  \zpm=\pm i \frac{\epm(z=0)}{\hpm(z=0)}~.
\eeq

\begin{figure}
\begin{center}
\includegraphics[width=0.4\textwidth]{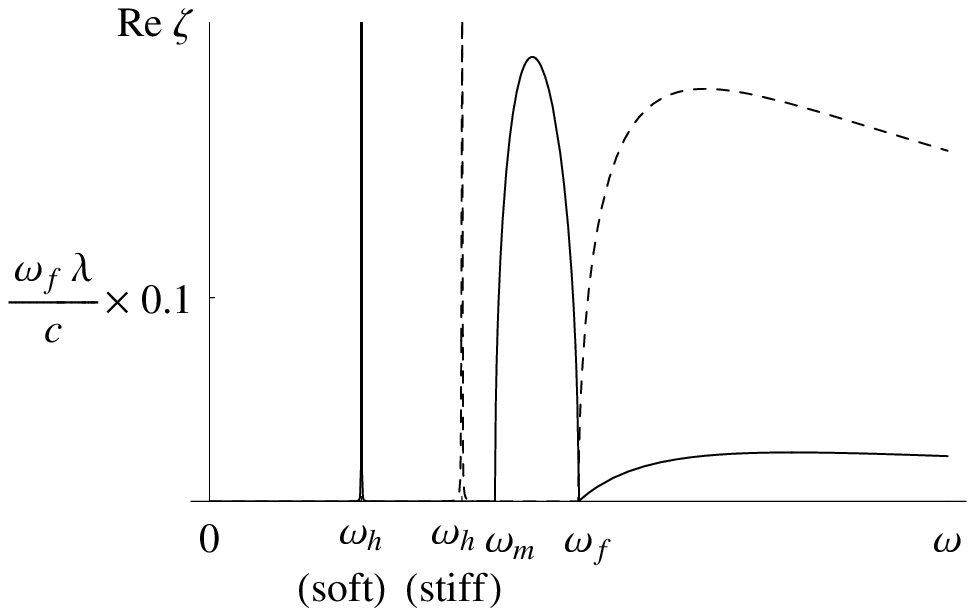}
\end{center}
\caption{$\re \,\z^\pp$ for the perpendicular geometry
 in the stiff regime, $\e=\sqrt{4\pi}+0.7$
(dashed line) and soft regime, $\e=\sqrt{4\pi}-1.7$  (solid line).
In both cases, $\a=4\pi+0.2$.} \label{zperp}
\end{figure}

Using Eqs.~(\ref{eq:he}) and (\ref{eq:perpBC}),
 we obtain, after simplification
\beq \label{eq:imp}
 \zpm=-
\frac{\o \l}{c}\frac{q_1 q_2 (q_1+q_2)}{1+q_1^2+q_2^2+q_1 q_2},
\eeq
where $q_i= k_i \l$.
Then, using the dispersion relation, Eq.~(\ref{eq:dispperp}), one can solve
for the wave vectors and obtain
a general expression for
the surface impedance as a function of the frequency:
\beq  \label{eq:imp2}
\zpm(\o)=-\frac{\o \l}{c} \frac{Q\sqrt{\pm \O-\d \a-\e^2+2\e Q}}{\pm\O-
\d \a+\e Q},
\eeq
where $Q= i\sqrt{\pm \O-\a}$.
It can be observed that $\z^\mm$ is pure imaginary since
spin waves with negative circular polarization cannot propagate, and it shows
regular behavior without special features. On the other hand, for the positive
polarization, the impedance has a real part satisfying the
inequality $\re\, \z^\pp
\ge 0$, as it should, and similar to the spectrum, it shows
a qualitatively different behavior in the two regimes depending on the value of
$\e$ as can be seen in Fig.~\ref{zperp}.
Most importantly, it has several singular points (thresholds) related to
spin-wave
propagation properties, which can be especially
useful for
comparison with experiments. Below we list these points and analyze the behavior
of $\z^\pp$ around them.

(i) The case $\e>\sqrt{4\pi}$ (stiff spin system). Then  $\O(k)$
is a growing function and the threshold  frequency for spin-wave
propagation is $\O_{f}=\a$. This frequency corresponds to uniform
oscillations in SCFM, at which the electric field vanishes inside
the sample.
 Near $\O_f$ one wave vector is small ($k_1 \propto\sqrt{\O-\O_f}$), and then
the surface impedance  can be simplified:

\beq
  \z^\pp\simeq \frac{\o_f \l}{c}\frac{\sqrt{(\e^2-4\pi)(\O-\O_f)}}{4\pi}.
      \label{eq:imp-stiff}
\eeq
Thus, $\re \z^\pp$ is zero below the threshold $\O_f$ and has a square-root
behavior above it

(ii) The case $\e<\sqrt{4\pi}$ (soft spin system). In this regime
the threshold frequency has a lower  value
$\O_m=\a-(\sqrt{4\pi}-\e)^2<\O_f$. At this frequency the electric
field on the sample surface vanishes (but unlike at $\O_f$, it is
nonzero inside the sample).
 Near $\O_m$, the two wave vectors  are
close in absolute value to $k_m$, but oppositely directed: $
k_1+k_2\propto \sqrt{\O-\O_m}$. Then the impedance is
\beq
  \z^\pp\simeq
  \frac{\o_m \l}{c \,\e} \sqrt{\frac{\sqrt{4\pi}-\e}{\sqrt{4\pi}}}
  \sqrt{\O-\O_m}.
\eeq
Thus, $\re \z^\pp$ has a square-root singularity at $\O_m$.

In addition, there is still a singularity at $\O_f$. The behavior
there, however, is different than in the stiff case: namely,
$\re\, \z^\pp$ is nonzero both above and below $\O_f$. In order to
obtain it,  Eq.~(\ref{eq:imp-stiff}) is not sufficient, and  one
has to go to the next order of smallness. Then
\beq
  \z^\pp\simeq \frac{\o_f \l}{c}\frac{\sqrt{4\pi-\e^2}}{4\pi}\lp
  [
  \frac{\e^3(\O-\O_f)}{4\pi(4\pi-\e^2)}  -i\sqrt{\O-\O_f} \rp],
\eeq
so for a soft spin system,  $\re\,\z^\pp$ vanishes at $\O_f$, showing a
square-root behavior below it and a linear one above it.

In addition to these points, where the impedance vanishes, there is a frequency
$\O_h= \d\a-\e^2/2+\e\sqrt{4\pi+\e^2/4}$
at which  $\z^\pp$ has a pole for both stiff and soft regimes:
\beq
     \z^\pp \simeq \frac{i \o_h \l}{c}\frac{(\a-\O_h)\sqrt{\e^2+
     \e\sqrt{\a-\O_h}}}{(\sqrt{\a-\O_h}+\e/2)(\O-\O_h)}.
\eeq Then $\re\, \z^\pp $ has a $\d$-function peak at $\o_h$ which
acquires a finite width if small dissipation is added to the
system. This peak corresponds to the ferromagnetic resonance for a
SCFM slab, and it
 is also related
to the surface wave supported by the system, whose branch starts
at the same frequency $\o_h$ in the limit of infinite light
velocity. \cite{SW} Indeed, the oscillation excited at this
frequency by an incident radiation is identical to the oscillation
produced in the slab by the long-wave surface wave.
 Finally, at
large frequencies $\O>>\a$ the magnetic properties stop playing a
role and the surface impedance becomes just the impedance of a
plain superconductor, $\z=-i\o \l/c$, with the real part being a
small correction $\sim \o^{-3/2}$.

Our analysis can be easily generalized to take into account
dissipation due to normal currents. For this,  normal current
$J_n=\s_n E$ should be added to the superconducting current in the
Maxwell equations. Here $\s_n$ is the normal conductivity and the
frequencies are assumed to be not very high, so the dispersion of
the conductivity can be neglected. As a result, in the London
equation (\ref{eq:london}), a renormalized complex $\l_n$ should
be used instead of the London penetration depth: \cite{clem}
 \beq
\label{eq:ln}
 \l_n^{-2}=\l^{-2}-2i \d^{-2},
\eeq where $\d=c/\sqrt{2 \pi \s_n \o}$ is the skin depth.
Consequently, the spin-wave spectrum and the surface impedance are
obtained from Eqs.~(\ref{eq:dispperp}), (\ref{eq:imp}), and
(\ref{eq:imp2}) by analytic continuation using complex $\l_n$ and
$\e$. Then, in general, the distinction between the soft and stiff
cases disappears, as $\re \,\z^\pp$ loses its singularities and
becomes nonzero at all frequencies, except at $\o_f$, where it
shows a square-root behavior both above and below $\o_f$. However,
if the normal currents are small, as happens in a superconductor,
 $\z$ behaves approximately in
the same way as in the nondissipative situation discussed above,
with the singularities smeared by a small amount
$\sim\l^2\d^{-2}$. In particular, the peak at $\o_h$ acquires a
finite width and can be observed.

The opposite limit  $\l \rightarrow \infty$ and $ \l_n^{2}=i
\d^{2}/2$ corresponds to a metallic ferromagnet. In this case $\re
\, \z$ again shows no features (except vanishing at $\o_f$). If,
however, the conductivity is not very high, $\g<<\d$, then the
frequencies $\o_h$ and $\o_m$ (calculated with a complex
$\e=\g/\l_n$) merge and the impedance shows a sharp peak there
(with width $\sim \g/\d$). This property has been widely used in
ferromagnetic resonance measurements in metals. \cite{akhi, Ament}
Thus, superconductivity modifies the microwave response in a
qualitative way, which can be used for detection of the
superconducting transition.

Finally, we would like to make the following comment. By analogy with metals,
one might think that the limit $\l \to \infty$ corresponds to an insulating
ferromagnet. This is not so, since in the description of the latter retardation
effects are important and should be taken into account. Accordingly, the correct
limit describing a magnetic insulator is $\l^{-2} \to -(\o/c)^2$.

\subsection{Finite applied field}
When a finite magnetic field $B_0<<H_{c2}$ is applied perpendicular to
the sample surface, it penetrates the sample, where the vortex phase is formed.
In the presence of vortices the London equation is modified since
the equation for the superconducting current density $\bs{j}_s$
now becomes
\beq
  \frac{4 \pi \l^2}{c} \nb \tm \sjs=B_0 \frac{\pd \su}{\pd z}-\sb,
\eeq
where $\su$ is the local displacement of the vortices from
the equilibrium position. To complete the description of the
problem in this regime, one needs also an equation for the vortex
dynamics. The simplest reasonable model is that the vortex motion
is governed by a balance of the Lorentz force and a viscous drag.
In the presence of normal currents, one should be careful about
the form of the Lorentz force, so that the Onsager relations are
not violated. \cite{Placais} A plausible assumption is that the
Lorentz force depends only on the superconducting current density.
With all these considerations taken into account, the vortex
dynamics is described by \beq
  \eta \frac{\pd \su}{\pd t}=\frac{\Phi_0}{c} \sjs \tm \bz
\eeq where $\eta$ is the viscous drag coefficient. Substituting
this equation in the previous one, we obtain \beq
  \frac{4 \pi \l_v^2}{c}\, \bz \tm \frac{\pd \sjs}{\pd z}=-\sb,
\eeq where $\l_v^2 = \l^2+ i B_0 \Phi_0/4 \pi \o \eta$. This
equation formally coincides with the London equation in the Meissner state,
 so in our approximations
the presence of vortices amounts to a replacement of the London
penetration depth $\l$ by the complex $\l_v$. Then the presence of normal
currents is taken into account by a further replacement \cite{Placais}
$\l_v^{-2}
\to \l_{nv}^{-2} = \l_v^{-2}-2i\d^{-2}$.

In addition, it is evident from Eq.~(\ref{eq:dm}) that instead of
the bare anisotropy parameter $\a$ a modified $\ta=\a+B_0/4\pi
M_0$ should be used. Thus, the effect of the applied field is: (i)
to shift the position of singularities in the response (this can
be useful since it allows to take measurements at  constant
frequency by scanning the applied field) and (ii) to add
dissipation to the system, smearing the singularities and
eventually pushing the response into the metallic regime.

\section{Parallel geometry}
\begin{figure}
\begin{center}
\includegraphics[width=0.4\textwidth]{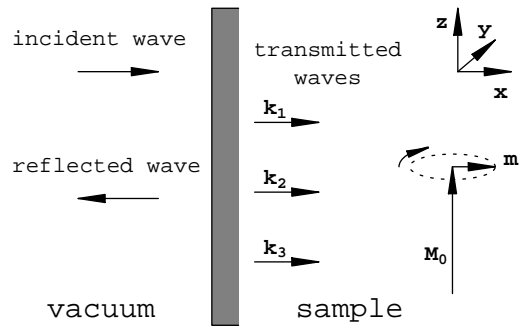}
\end{center}
\caption{The setup in the parallel geometry.}
\label{par}
\end{figure}

Now we consider a situation when the easy axis is parallel to the
sample surface, as shown in Fig.~\ref{par}. The direction
perpendicular to the surface is chosen to be $\bx$, along which
all quantities vary, and then $\smp=\by \, m_y$. Again we start
from a simpler case when the magnetic induction vanishes inside
the sample, $B_0=0$. After that, we consider a more complicated
case when there is a nonzero induction near the sample surface,
decaying in the bulk.

\subsection{Zero static induction}
The regime in which the static magnetic induction vanishes inside
the sample can be realized by application of an external field
$\bho=-4\pi \bmo$. Strictly speaking, this configuration is not stable,
since flipping $\bmo$ in the opposite direction would lower the energy.
However, it  is
 metastable provided the anisotropy is strong enough, $\a>4\pi$. From
Eq.~(\ref{eq:B0=0}) the spectrum of spin waves now is
\beq
\label{eq:disppar}
  \O^2=\lp(\a+\g^2 k^2\rp)\lp(\a+\g^2 k^2-\frac{4\pi k^2 \l^2}{1+k^2 \l^2}\rp).
\eeq
It can be seen from this relation  that
 in the parallel
geometry an incident electromagnetic wave excites in the sample
three spin-wave modes. Since the dynamical part of the
magnetization has only one component in the plane of the sample
surface, $m_y$, the spin wave in this geometry is excited by a
linearly polarized incident radiation with the magnetic component
in the $\by$ direction. The spectrum   has the same qualitative
features as in the perpendicular geometry, although the specific
values of different parameters are, of course, different. Thus,
the transition between the soft and stiff regimes occurs now at
$\e=\sqrt{2\pi}$, while the wave vector $\kmp$ and the frequency
$\omp$ can be found by minimizing Eq.~(\ref{eq:disppar}), which
requires solution of a cubic equation. Note, however, that the
frequency of uniform oscillations $\O_f$ remains equal $\a$  in
this geometry.

The
boundary condition, Eq.~(\ref{eq:perpBC}), now gives two independent
equations for
$\bx$ and
$\by$ components of the magnetization. These two components are related by
Eq.~(\ref{eq:B0=0}):
\beq \label{eq:mxmy}
  i \O \, m_y=(\a+\g^2 k^2)m_x.
\eeq
Using this, the boundary conditions can be written as
\bear \label{eq:BCpar}
 && \sum k_i \,m_{x \,i}=0 \nonumber \\
 && \sum k_i^3 \, m_{x \,i}=0 ,
\eear
where the summation is over the three spin-wave modes. These conditions
determine the relative amplitudes of the modes and together with the dispersion
relation specify the problem completely.

\begin{figure}
\begin{center}
\includegraphics[width=0.4\textwidth]{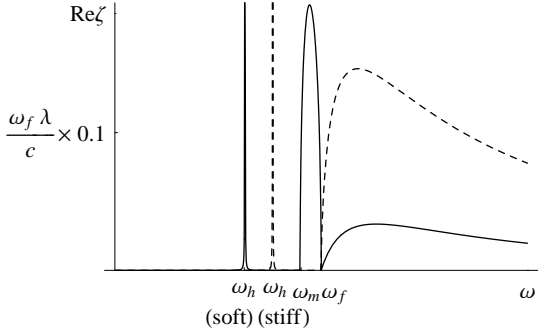}
\end{center}
\caption{$\re \,\z$ for the parallel geometry
 in the stiff regime, $\e=\sqrt{2\pi}+0.1$
(dashed line) and soft regime, $\e=\sqrt{2\pi}-1.$  (solid line).
In both cases, $\a=4\pi+0.3$.} \label{zpar}
\end{figure}

The surface impedance for this case of linearly polarized incident wave is
given by
\beq
  \z=-\frac{e_z(x=0)}{h_y(x=0)}.
\eeq Then the calculation goes along the same steps as for the
perpendicular geometry, though it is more complicated due to the
presence of three modes. Using Eqs.~(\ref{eq:he}),
(\ref{eq:mxmy}), and (\ref{eq:BCpar}), one obtains, after some
algebra,
\beq \label{eq:zetapar}
  \z=\frac{\o \l}{c}\,\frac{a_2(a_1 a_3-a_2)\e^4}{a_1(\O^2-\a\,\d\a)+
  a_2[(2\a-4\pi)\e^2-a_3 \e^4]},
\eeq
where $a_j$ are symmetric combinations of the normalized wave vectors for
different
modes (and, as before, $q_i=k_i \l$):
\bear \label{eq:a}
  a_1&=&q_1+q_2+q_3, \nonumber \\
  a_2 &=&q_1 q_2 q_3,\nonumber \\
  a_3 &=&q_1 q_2 +q_2 q_3+q_3 q_1
.\eear
These combinations can be found from the dispersion relation,
Eq.~(\ref{eq:disppar}), thus providing the dependence of $\z$ on the frequency.
As can be seen from Fig.~\ref{zpar}, $\re \,\z$ shows the same
qualitative behavior as in the perpendicular geometry around its singularities.
The same is also true about the high-frequency behavior of $\z$, given by
$-i \o \l/c$ with a small real correction $\sim \o^{-3/2}$.
Below we give expressions for $\z$ near the propagation thresholds.

(i) In the stiff regime $\e>\sqrt{2\pi}$, the impedance near the
threshold $\O_f=\a$ shows a square-root singularity \beq
  \z\simeq  \frac{\o_f \l}{ c}\frac{\sqrt{(\e^2-2\pi)(\O-\O_f)}}{2\pi}
\eeq

(ii) In the soft regime $\e<\sqrt{2\pi}$, the threshold is at
$\omp$, again with a square-root behavior around it: \bear
\label{eq:parsoft}
  \z&\simeq & \frac{\omp \l}{c} \frac{\qmp(\a-\e^2)\sqrt{3\e^2 \qmp^2+2\a-4\pi
  +\e^2}}{2\e (\a+\e^2 \qmp^2)(\e^2\qmp^2+\a-2\pi)\sqrt{1+\qmp^2}}
  \nonumber \\ &&
 \tm  \sqrt{\O^2-\Omp^2}.
\eear
Finally, Eq.~(\ref{eq:zetapar}) has a pole at a frequency $\ohp$, at which
$\re \z$ shows a peak, just like in the perpendicular geometry.

\subsection{Finite static induction}
If the applied magnetic field  $\bho$(parallel to the surface) is
not exactly equal to $-4\pi \bmo$, then there is a nonzero
magnetic induction penetrating the sample, \beq \label{eq:b}
  \bbo(x)=(4\pi \bmo+\bho)
  e^{-x/\l}.
\eeq In this case the translational invariance of the system is
broken, so the spin-wave modes are not plane waves anymore. This
considerably complicates the problem, since these modes are given,
instead of Eq.~(\ref{eq:disppar}), by a sixth-order differential
equation \bear \label{eq:waveB}
  &&\O^2(\l^{-2}-\pd_x^2)m_x=(\l^{-2}-\pd_x^2)(\a+f e^{-x/\l}-
  \g^2 \pd_x^2)^2 m_x
  \nonumber \\ &&+
  4\pi \pd_x^2 (\a+f e^{-x/\l}-\g^2 \pd_x^2)m_x,
\eear where $f=(4\pi+H_0/M_0)$ is the magnetic induction at the
sample surface divided by $M_0$. Then the general solution of the
problem including finding the modes, matching them by the boundary
conditions, and calculating the surface impedance would be very
difficult. Instead of this we limited our attention to two extreme
regimes in which the problem can be simplified. Namely, we
considered the regime of very low stiffness (very soft regime),
$\e<<1$, and the opposite very stiff regime $\e>>1$. In the former
case,
 the Meissner layer
is wide and the static magnetic field there is very smooth, and
therefore the quasiclassical approximation can be applied; in the
latter, the Meissner layer is extremely thin and its influence is
weak.

\subsubsection{Very soft regime}
In this regime $\e<<1$, and the dispersion of free spin-wave modes (those outside of the
Meissner layer) can be obtained by simplification of Eq.~(\ref{eq:disppar}):
\beq \label{eq:freespect}
  \O^2=(\a+\g^2 k^2)(\d \a+\g^2 k^2)+\frac{4\pi}{1+k^2 \l^2}.
\eeq
Then the threshold frequency $\Omp$ for spin-wave penetration is given by
\beq \label{eq:omp}
 \Omp^2=\a \d\a+
4\e\sqrt{2\pi \a(\a-2\pi)}+ O(\e^2)
 \eeq
 and the corresponding wave vector is
\beq k_m=\frac{1}{\sqrt{\g\l}}\sqrt[4]{\frac{2\pi \a}{\a-2\pi}}.
\eeq Of the three spin-wave modes corresponding to a general
frequency $\O$, two have short wavelength $\sim \g$, while one has
long wavelength $\l$. However, for frequencies near $\Omp$, the
picture changes: namely, one has a short-wave decaying mode with
$k_3^2=-2(\a-2\pi)\g^{-2}$ and two modes with intermediate
wavelengths $\sim \sqrt {\g\l}$. Since in this case all
wavelengths are much shorter than the scale of variation of the
``potential'' [provided by the field $\bbo(x)$], it is possible to
treat the spin-wave equation (\ref{eq:waveB}) by the
quasiclassical approximation (its validity is discussed below in
more detail).

Thus we will be interested in finding $\z$ (and in particular, its real part)
for frequencies slightly above the threshold, $0<\O-\Omp\sim \e$ and in the presence
of weak magnetic induction $B_0\sim \e M_0$ directed either parallel or
antiparallel to $\bmo$. We start by discussing $\z$ in the absence of $B_0$.
From the free-wave spectrum, Eq.~(\ref{eq:freespect}), one finds relations
between the wave vectors $k_1$ and $k_2$:
\bear \label{eq:vsoft12}
   k_1^2+k_2^2=\frac{\O^2-\a \d\a}{2(\a-2\pi)}\g^{-2},
   \nonumber \\
   k_1^2 k_2^2 =\frac{2\pi \a}{\a-2\pi}\g^{-4} \l^{-2},
\eear
while the short-wave-mode wave vector $k_3$ has been specified earlier.
From these
relations it is not difficult to find the wave vectors and substitute them
into Eq.~
(\ref{eq:zetapar}), thus obtaining
\beq \label{eq:vsoft}
   \z=\frac{\omp \l}{c} \frac{\sqrt{4\pi \a}\sqrt{\O^2-\Omp^2}}{\O^2-\Ohp^2},
\eeq
where the resonance frequency $\Ohp$ is given (in the very soft regime) by
\beq
    \Ohp^2=\a \d\a+2\e\sqrt{2\pi\a(\a-2\pi)}.
\eeq We recall that Eq.~(\ref{eq:vsoft}) is valid only near
$\omp$, where Eqs.~(\ref{eq:vsoft12}) hold. To lowest order in
$\o-\omp$, Eq.~(\ref{eq:vsoft}) reduces to \beq
  \z=\frac{\omp \l}{c} \frac{\sqrt{\Omp(\O-\Omp)}}{\e\sqrt{\a-2\pi}},
\eeq
which coincides with Eq.~(\ref{eq:parsoft}) in the very soft ($\e<<1$) limit.

\begin{figure}
\begin{center}
\includegraphics[width=0.4\textwidth]{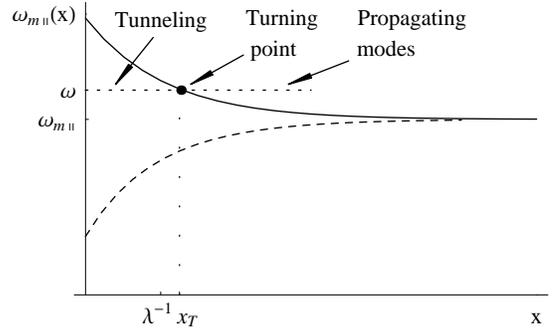}
\end{center}
\caption{Coordinate-dependent threshold frequency $\omp(x)$
serving as an effective potential for $B_0>0$ (solid line) and
$B_0<0$ (dashed line).} \label{quasicl}
\end{figure}

Let us now turn on the magnetic induction $\bbo(x)$; then, its
presence can be effectively described by an $x$-dependent
anisotropy factor $\a(x)= \a+B_0(x)/M_0$. According to the
quasiclassical approximation, the corresponding spin-wave modes
are found from the free modes by using $x$-dependent wave vectors.
Note that for the  short-wave mode 3 this dependence can be
neglected since it is small and regular. On the other hand, for
the modes 1 and 2, this dependence is crucial, since it determines
properties of the wave propagation such as tunnelling. Indeed,
substituting $\a(x)$ into Eq.~(\ref{eq:omp}), one obtains the
$x$-dependent threshold frequency \beq
    \Omp^2(x)=\Omp^2+2(\a-2\pi)f e^{-x/\l},
\eeq
 which can be considered as an effective
 potential as shown in Fig.~\ref{quasicl}. In the region where
 $\O>\Omp(x)$, modes 1 and 2 are propagating,
 while if $\O<\Omp(x)$, these modes can only tunnel under the potential
barrier. The point $x_T$ between these two regions is a classical turning, or
reflection point. It is defined by the condition
\beq \label{eq:xt}
  \O=\Omp(x_T).
\eeq
 The field-free threshold $\Omp=\Omp(x=\infty)$ remains
the propagation threshold so that only radiation with $\O>\Omp$
can penetrate (here we ignore the peak at $\ohp$). When $B(x)<0$,
this penetration is classically possible, since $\O>\Omp(x)$,
while when $B(x)$ is positive and not too small, so that
$\Omp(0)>\O$, it is realized by tunneling.

The validity of the quasiclassical approximation is of special concern in this
problem. Indeed, as was mentioned above, one can hope to use the quasiclassics
only near the threshold $\Omp$. On the other hand, $\Omp(x)=\O$ means that
the motion is near the turning point, around which the quasiclassics breaks down.
The usual condition of validity
$\pd_x k(x)<<k^2$, is not sufficient here, since, unlike in quantum mechanics,
 the turning point corresponds not to zero but a finite wave vector $k_m$ so that
 $\O(k)-\Omp \propto (k-k_m)^2$ (for the mode with $\re \, k>0$). Instead,
 a more stringent condition is required,
 \beq \label{eq:kcond}
   \frac{\pd \, \d k}{\pd x}<< \d k^2,
 \eeq
where $\d k = k-k_m$. There is also  another dangerous frequency
given by $\O_-^2=\a \d\a-4\e\sqrt{2\pi \a(\a-2\pi)}$, at which
$k^2=-k_m^2$, the group velocity vanishes and the quasiclassics
breaks down again (this frequency has no quantum-mechanical
counterpart).
 However, it is situated deep inside the classically forbidden region, and we
 ignore
it, assuming $\O$ to be always larger than it. Then the above
condition reduces to
\beq \label{eq:cond}
   \e f e^{-x/\l}<<\lp|\frac{\O^2-\Omp^2(x)}{\a-2\pi}\rp|^{3/2}.
\eeq

(i) For $B(x)<0$, the effective potential bends down. Then, as
long as the quasiclassics holds, the wave propagation inside is
described by two outgoing modes 1 and 2 (plus the evanescent
short-wave mode 3) with $x$-dependent wave vectors found from
Eq.~(\ref{eq:vsoft12}). Then $\z$ can be calculated exactly as in
the field-free case, but with a local value of $\a$ at the
surface, $\a(x=0)=\a+f$. The result is \beq \label{eq:overbar}
   \z=\frac{\omp \l}{c} \frac{\sqrt{4\pi \a}\sqrt{\O^2-\Omp^2-2(\a-2\pi)f}}
   {\O^2-
   \Ohp^2-2(\a-2\pi)f},
\eeq which has just the field-free form shifted to the left
($f<0$). One might worry about this result that it gives real $\z$
for $\O<\Omp$. In fact, as $\O \to \Omp$, the quasiclassics breaks
down at some depth inside the slab, and this result loses its
validity. This is expressed in the appearance of significant
over-barrier reflection. Then the correct frequency dependence
becomes $\re \z \propto \sqrt{\O-\Omp}$. However, as long as $\O$
is not too close to $\Omp$ [e.g., for $f \sim \O-\Omp \sim \e$,
when condition~(\ref{eq:cond}) is satisfied for any $x$], the
result given above is valid.

\begin{figure}
\begin{center}
\includegraphics[width=0.4\textwidth]{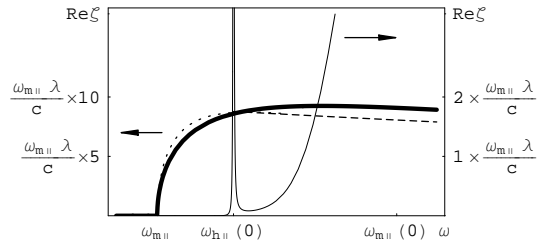}
\end{center}
\caption{Effect of magnetic field in the parallel geometry, very
soft regime. $\re \,\z$ for $f=0$ (thick line), $f=0.05$ (thin
solid line, its scale given to the right), and $f=-0.05$ (dashed
line), which is extrapolated to have a square-root behavior near
$\o_m$ (dotted line). Note that the values of $\ohp(0)$ and
$\omp(0)$ are given for a positive field $(f=0.05)$. Parameter
values for the plots are $\a=4\pi+1$ and $\e=0.01$.}
\label{bplots}
\end{figure}

(i) For $B(x)>0$, the effective potential bends up, forming a
barrier. For frequencies $\Omp<\O<\Omp(0)$, the wave penetration
is realized by tunneling of the modes 1 and 2, which means that
each of these modes consists of a dominant decreasing part and a
small (subdominant) increasing part (while the mode 3 remains
unaffected). The tunneling occurs until the turning point $x_T$,
after which the modes become propagating. In order to calculate
$\im \,\z$, one may neglect the tunneling, thus assuming that all
the energy  is reflected from the barrier. Then only the dominant
part of the modes should be retained, the problem becomes again
formally analogous to the free case, and $\im \,\z$ is given by
Eq.~(\ref{eq:overbar}) with $f>0$. Calculation of $\re\, \z$ is
more difficult, since it requires higher precision. It can be done
directly by keeping both the dominant and subdominant parts and
solving a boundary-value problem with five components. However, it
is easier to solve the problem using the fact that $\re \z$ is
proportional to the energy flux: \beq \label{eq:tunnel}
   \re \,\z =\frac{8\pi}{c}P|h_y(0)|^{-2}.
\eeq
The flux may be calculated in the region after the turning point, where it
is given by
\beq \label{eq:flux}
  P=\sum_{i=1,2} \frac{\o_m |m_{x\,i}(x)|^2}{4\a}\frac{\pd \O^2(k[x])}{\pd k},
\eeq
with $\O(k)$  given by
Eq.~(\ref{eq:freespect}). The transmitted amplitudes may be related to the
amplitudes at the boundary by a
 procedure analogous to the quasiclassical
calculation of  transmission coefficient in quantum mechanics,
\cite{LL3} neglecting the subdominant parts. When this is done,
one has to relate different mode amplitudes at $x=0$ using the
boundary conditions, Eq.~(\ref{eq:BCpar}). Substituting the
results into Eqs.~(\ref{eq:tunnel}) and (\ref{eq:flux}), one
obtains
\beq
  \re \,\z=\frac{8\l\,\omp}{c}\frac{\e^{3/2} (\a-2\pi)^{3/4}(2\pi \a)^{5/4}\sqrt{
  \Omp^2(0)-\O^2}}{[\O^2-\Ohp^2(0)]^2\sqrt{\O^2-\O_-^2(0)}}\,t,
\eeq where for all threshold frequencies, $\O^2(0)=
\O^2+2(\a-2\pi)f$ and \beq t= \exp\lp[-2\int_0^{x_T} \im k(x) \,
dx \rp] \eeq is the tunneling exponent. Using
Eqs.~(\ref{eq:vsoft12}) and (\ref{eq:xt}), one finds
\beq
t=\exp\lp[-2\e^{-1}\sqrt{f e^{-\xi_T}} \lp(\sqrt{e^{\xi_T}-1}-
\arctan\sqrt{e^{\xi_T}-1} \rp) \rp],
 \eeq
 with
 \beq
  e^{\xi_T}= e^{x_T/\l}=2f\frac{\a-2\pi}{\O^2-\Omp^2}.
\eeq
Note that $\z$ may have a peak of finite width if the field is strong enough
so that $\ohp(0)
>\omp$. The peak is cut off at $\O^2-\Ohp^2(0)\sim \e^{3/2}t$, when the
subdominant part becomes significant and the tunneling is no
longer small, as was assumed. On the other hand, the singularity
at $\O_-(0)$ is disregarded, since, as we mentioned above, this
frequency is assumed to be always smaller than $\Omp$. The
calculation remains valid for $\O$ going down to $\Omp$. The
reason for this is that unlike in the previous case, the
quasiclassics now does not need to hold everywhere in the sample
(it is anyway broken around the turning point), but only in some
finite region under the barrier and deep inside after $x_T$. It
can be seen from Eq.~(\ref{eq:cond}) that these requirements are
satisfied for $0<\O-\Omp \lesssim f \sim \e$. Plots of $\re \,\z$
for $B_0=0, B_0<0,$ and $B_0>0$ are shown on Fig.~\ref{bplots}.
One can see that for the antiparallel field ($B_0<0$) the response
is not affected significantly, unlike for the parallel field.

\subsubsection{Very stiff regime}
In this regime $\e>>1$, which corresponds to the limit of very strong
superconductivity. Thus the Meissner layer is very thin, and its
influence amounts to  modification of the boundary conditions.
 The threshold for wave
propagation in this regime is $\O_f=\a$.
Again we start  by
considering the field-free situation first. The free spin-wave spectrum,
Eq.~(\ref{eq:disppar}), yields in this regime one short-wave and two long-wave
modes:
\beq
  k_1^2\approx -\l^{-2}, \quad  k_{2,3}^2\approx \g^{-2}(\pm\O-\a).
\eeq Substituting them into Eqs.~(\ref{eq:zetapar}) and
(\ref{eq:a}), one finds the field-free result
\beq
   \z=\frac{\o \l}{c}\lp(-i+\frac{2\pi}{\e\sqrt{\O-\a}}\rp).
\eeq Thus to the leading order $\z$ is pure imaginary and
corresponds to the surface impedance of a usual nonmagnetic
superconductor, as it should in this regime of strong
superconductivity. Note that this expression is not valid very
near the threshold - namely, for $\O-\a\lesssim \e^{-2}$ - and
hence it does not capture the special behavior there, such as the
square-root singularity $\z\sim\sqrt{\O-\a}$ and a pole at $\Ohp$
(in this regime $\ohp$ and $\o_f$ are close to each other).

Now we introduce the magnetic field, Eq.~(\ref{eq:b}), which is not assumed
 small,
 and
calculate the change in $\re \,\z$ due to it. As will be shown
below, the change comes only as a small correction $\sim \e^{-2}$
to the field-free results; hence, small terms $\sim \e^{-2}$
should be retained in the calculation. To start with, it is useful
to rewrite the spin-wave equation, Eq.~(\ref{eq:waveB}), in terms
of a rescaled coordinate $\xi = x/\l$:
\bear
  \O^2(1-\pd_\xi^2)m_x=(1-\pd_\xi^2)(\a+f e^{-\xi}-\e^2\pd_\xi^2)^2 m_x
  \nonumber \\+
  4\pi\pd_\xi^2 (\a+f e^{-\xi}-\e^2 \pd_\xi^2)m_x.
\eear
This equation contains a large parameter $\e$, which makes it possible to find
the solutions as an expansion in $\e^{-1}$. As in the field-free case, one
has  a short-wave mode
\beq
    m_{x\,1}(\xi)\approx m_1 e^{-\xi}\lp(1+\frac{5 f}{16}\e^{-2} e^{-\xi} \rp)
\eeq
and two long-wave modes
\beq
  m_{x \,\,2,3}(\xi) \approx m_{2,3} \exp(i q_{2,3} \xi) (1+f \e^{-2}
  e^{-\xi}),
\eeq where we have only retained corrections related to the field
($\sim f$). Similarly one finds the modes for $m_y$,
 \bear
   m_{y\,1}(\xi)=-\e^2\frac{m_1 e^{-\xi}}{i\O}\lp(1+\frac{f \e^{-2}e^{-\xi}}{4}
   \rp),
   \nonumber \\
   m_{y\,\,2,3}(\xi)=\mp i m_{2,3} \exp(iq_{2,3} \xi)(1+\e^{-2}f e^{-\xi}),
\eear where in the second equation the upper sign is used for mode
2 and lower for mode 3, and analogously for
 $b_y$,
 \bear
 && b_{y\,1}(\xi)=\e^4\frac{m_1 e^{-\xi}}{i\O},
  \nonumber \\
&&  b_{y\,2,3}(\xi)=\mp 4\pi i m_{2,3}
  \exp(iq_{2,3} \xi)\lp(
  q_{2,3}^2-\frac{\e^{-2}f e^{-\xi}}{2iq_{2,3}}\rp),\quad
 \eear
and $h_y$,
 \bear \label{eq:vstiffH}
 && h_{y\,1}(\xi)=\frac{\e^4 m_1 e^{-\xi}}{i\O}   \nonumber \\
&&  h_{y\,2,3}(\xi)=\pm 4\pi i m_{2,3}\exp(iq_{2,3} \xi)\lp(1+\frac{\e^{-2}f
  e^{-\xi}}{2iq_{2,3}}\rp).\quad
\eear Now one should use the boundary conditions in order to
relate the amplitudes of different modes. It would be incorrect to
use the conditions $\nabla \sm(0)$ directly since differentiation
would reduce the relative precision in the long-wave modes and the
result would only have a precision of $\e^{-1}$. Instead, one
should integrate the equations for magnetization components,
Eq.~(\ref{eq:dm}), over the Meissner layer and use the above
condition in the integral of $\nabla^2 \sm$, leading to
\bear
&&\int [f e^{-\xi}m_y(\xi)- b_y(\xi)]\, \rd x=\e^2 (q_2 m_2-q_3 m_3),
 \nonumber\\
&&\int  f e^{-\xi}m_x(\xi)\, \rd x=\e^2(iq_2 m_2+iq_3 m_3).
\eear
Substituting the expressions for the magnetization components and $b_y$, one
 obtains relations between the mode amplitudes:
\bear
 &&   \frac{\e^2}{\O}m_1\lp(1+\frac{\e^{-2}f}{2}\rp)=
    \nonumber \\
&&  \quad  \sum_{j=2,3} \pm m_j \lp(-i q_j+\e^{-2}f(1+i q_j)
    +2\pi\frac{\e^{-4}f}{i q_j} \rp)
    \nonumber \\
&&    m_2(i q_2-\e^{-2}f)+m_3(i q_3-\e^{-2}f)=0,
 \eear
 where again
the upper and lower signs in the first equation refer to modes 2
and 3, respectively. These relations are used in  an expression
for the energy flux in the region deeper than the Meissner layer,
\beq
  P=\o\l \e^2 |m_2|^2 \re\, q_2
\eeq
(where only mode 2 contributes  as it is the only propagating
mode), and in   Eq.~(\ref{eq:vstiffH}), which are subsequently substituted in
Eq.~(\ref{eq:tunnel}) to calculate $\re\, \z$. The result is
\beq
  \re \,\z= \frac{\o \l}{c} \frac{2\pi \e^{-1}}{\sqrt{\O-\a}} \lp(1+3\e^{-2}f+
  \frac{8\pi \a\e^{-2}f}{\O^2-\a^2}-\frac{f^2 \e^{-2}}{\O-\a} \rp).
\eeq Thus the influence of the field is given by a sum of three
contributions which can be extracted by considering the frequency
and field dependences of the response. All these contributions
come about as small corrections $\sim \e^{-2}$, although the field
itself (near the surface) is not small. This is because the
Meissner layer is very thin.

\section{Conclusions}
 I have considered the microwave response properties of a slab made
of a material with coexisting superconductivity and magnetism and
calculated its surface impedance for the parallel and
perpendicular geometry. The impedance, and especially its real
part, shows a rich behavior around threshold frequencies, related
to spin-wave propagation. In the stiff regime, the propagation
threshold is $\o_f$, above which $\re \, \z$ shows a square-root
singularity. On the other hand, in the soft regime this behavior
is observed near a lower threshold $\o_m$, while around $\o_f$,
$\re \, \z$ has a square-root singularity below it and a linear
behavior above it. In addition, there is a resonance peak at
$\o_h$ in both regimes, related to excitation of a surface spin
wave. \cite{SW} The high-frequency response is mostly imaginary,
being dominated by the superconductivity, with a small real
correction $\sim \o^{-3/2}$.

This behavior is common for both geometries in the field-free
situation, when no magnetic induction exists at the sample
surface. However, when such an induction does exist, its influence
is completely different in these geometries. In the perpendicular
geometry, the magnetic induction penetrates the slab in form of
vortices. This leads to a change in the effective anisotropy
constant, resulting in a uniform shift of all thresholds and
allowing one to detect the resonances by sweeping the magnetic
field at constant frequency. However, it also leads to an
increased dissipation, smearing the singularities and eventually
pushing the response into the metallic regime.

On the other hand, in the parallel geometry, the magnetic induction is
screened inside the Meissner layer. We found the influence of the induction
in the regimes of  very low and very high magnetic stiffness. In the former limit the response near
$\omp$ is strongly direction-dependent: for $\bb$ and $\bmo$ in the opposite
directions, the response is not changed significantly, while if they are
parallel to each other, $\re \, \z$ is exponentially suppressed due to a
potential barrier formed in the Meissner layer. In addition, it  also
shows a singularity $\sim[\o-\ohp(0)]^{-2}$, once the induction is strong enough
so that $\ohp(0)>\omp$. In the opposite very stiff regime dominated by
 strong
superconductivity, the induction is screened out very fast and its
influence is  weak. Interestingly, the directional dependence of
the response is opposite to the previous case, so that $\re
\z(B_0>0)>\re \z(B_0<0)$.

These results demonstrate that the magnetic and superconducting
properties of SCFM's (including unconventional superconductors
with orbital magnetism) can be effectively investigated by
microwave response measurements and provide a theoretical basis
for such experiments.

\begin{acknowledgments}
I am grateful to E.~B.~Sonin for many illuminating discussions.
This work has been supported by the Israel Academy of Sciences and
Humanities.
\end{acknowledgments}

\end{document}